\documentclass[letterpaper]{article} 

\usepackage[]{aaai25}  
\nocopyright 
\usepackage{times}  
\usepackage{helvet}  
\usepackage{courier}  
\usepackage[hyphens]{url}  
\usepackage{graphicx} 
\usepackage{natbib}  
\usepackage{caption} 
\usepackage{subcaption} 
\usepackage{float}      
\usepackage{algorithm}
\usepackage{algorithmic}
\usepackage{booktabs}
\usepackage{amsmath}
\usepackage{accessibility} 

\usepackage{newfloat}
\usepackage{listings}
\DeclareCaptionStyle{ruled}{labelfont=normalfont,labelsep=colon,strut=off} 
\floatstyle{ruled}
\newfloat{listing}{tb}{lst}{}
\floatname{listing}{Listing}

\usepackage{xcolor}

\title{Structural Dynamics of Harmful Content Dissemination on WhatsApp}
\author{
Yuxin Liu\textsuperscript{\rm 1},
M. Amin Rahimian\textsuperscript{\rm 1,*},
Kiran Garimella\textsuperscript{\rm 2,*}
}

\affiliations{
\textsuperscript{\rm 1}University of Pittsburgh \\
\textsuperscript{\rm 2}Rutgers University \\

yul435@pitt.edu, rahimian@pitt.edu, kiran.garimella@rutgers.edu \\

$^*$ Corresponding authors.
}
\usepackage{bibentry}

\usepackage{xcolor}

\begin{document}







\maketitle

\begin{abstract}

WhatsApp, a platform with more than two billion global users, plays a crucial role in digital communication, but also serves as a vector for harmful content such as misinformation, hate speech, and political propaganda. This study examines the dynamics of harmful message dissemination in WhatsApp groups, with a focus on their structural characteristics. Using a comprehensive data set of more than 5.1 million messages–including text, images, and videos–collected from approximately 6,000 groups in India, we reconstruct message propagation cascades to analyze dissemination patterns.

Our findings reveal that harmful messages consistently achieve greater depth and breadth of dissemination compared to messages without harmful annotations, with videos and images emerging as the primary modes of dissemination. These results suggest a distinctive pattern of dissemination of harmful content. However, our analysis indicates that modality alone cannot fully account for the structural differences in propagation. 
The findings highlight the critical role of structural characteristics in the spread of these harmful messages, suggesting that strategies targeting structural characteristics of re-sharing could be crucial in managing the dissemination of such content on private messaging platforms.

\end{abstract}

\section{Introduction}
WhatsApp is the world’s most popular messaging app, with more than 2.78 billion monthly active users in 180 countries, and is especially dominant in countries like India. In fact, more than 500 million users in India rely on WhatsApp as their main means of communication, making it a central part of daily life for many~\cite{Statista2023}. Despite its vast user base, studying information dissemination on WhatsApp remains challenging due to its end-to-end encryption, which limits access to message content--even to the platform--and makes it difficult to monitor and track harmful material such as misinformation, hate speech, and propaganda. 
Due to the lack of moderation, malicious actors have exploited WhatsApp's group chat feature to disseminate propaganda and hateful content on a large scale~\cite{evangelista2019whatsapp,nizaruddin2021role}. 

Recent events have highlighted the severe real-world consequences of widely distributed harmful messages on WhatsApp. Since 2014, there has been a significant increase in mob violence, often triggered by misinformation spread through WhatsApp. For example, in India, false rumors about child kidnappers circulated on WhatsApp, leading to the deaths of more than two dozen innocent people since April that year~\cite{arun2019whatsapp}. 
Beyond violence, WhatsApp has also become a powerful tool for political manipulation. In India, where over 500 million people use WhatsApp, both the BJP\footnote{The Bharatiya Janata Party is the ruling party of India and largest in terms of political representation in the Parliament and state legislatures.} and the Congress\footnote{The Indian National Congress (INC), colloquially ``the Congress'', is one of the oldest political parties in India.} have been accused of spreading false or misleading information to influence the country's 900 million large voting population, highlighting the role of the platform in shaping public opinion~\cite{hindustantimesModis2019}.

Despite the urgency of understanding how harmful messages spread on WhatsApp, tracking their diffusion remains particularly challenging due to the end-to-end encryption of the platform. Unlike public platforms such as X (formerly Twitter) or Facebook, WhatsApp operates as a private communication network where messages are encrypted before sending and only decrypted on the receiver's device. This end-to-end encryption ensures that even the messaging platform itself cannot access the content exchanged between users. Consequently, developing machine learning algorithms to detect the spread of harmful or false information is virtually impossible, as the platform cannot analyze the messages in transit. Without access to message content, conventional content moderation techniques, as seen on public platforms, cannot be applied. This limitation makes it extremely difficult to monitor or control the dissemination of misinformation on WhatsApp. 

Given the impracticality of content-based moderation, WhatsApp has changed its focus toward the structural characteristics of message dissemination to limit the spread of harmful messages. A key strategy implemented by WhatsApp is to restrict the message forwarding functionality. Specifically, each message can be forwarded to a maximum of five groups, and if the message has already been forwarded multiple times, it can only be forwarded to a single group. Furthermore, once a message has been forwarded more than five times, it is marked with the label ``forwarded many times.'' However, it remains an open question whether harmful messages exhibit structural dissemination patterns that are distinct from normal content. This issue is critical, as the effectiveness of WhatsApp's structural restrictions relies on the assumption that harmful messages follow identifiable patterns, which can be targeted by these restrictions.

In this paper, we built a large dataset collected by~\citet{garimella2024whatsapp} comprising 5,158,879 messages, including text, images, and videos, collected from 5,953 WhatsApp groups, which are a key channel for the mass dissemination of information on the platform. This dataset was collected through a field study in India, where participants gave their informed consent to share data from the WhatsApp groups that they were comfortable with, ensuring ethical compliance. To track and analyze the spread of messages, Garimella and Chauchard used privacy-preserving hashing techniques to identify multiple instances of the same message, even with slight variations. Techniques such as PDQ hashing~\cite{farid2021overview} for images and videos and locality-sensitive hashing (LSH) ~\cite{gionis1999similarity} for text, allowed them to detect content variations, such as cropping or encoding changes that capture exact and modified versions of shared media.

We assigned labels for misinformation, hate speech, and propaganda to a small subset of 2,000 viral pieces of content in this dataset. Taking into account the data donation-based sampling strategy, we model the message cascades as generalized tree-like structures and estimate key structural parameters, namely breadth $b$ and depth $h$, to understand the dissemination process. 


Our analysis revealed that harmful messages have significantly greater breadth and depth compared to normal messages. In addition, these harmful messages are spread primarily through videos and images. We also estimated the scale of dissemination for different message types based on group sizes to derive population-level estimates and found that harmful messages reach a significantly larger audience compared to normal messages. These findings emphasize the need for further investigation into the structural aspects of message propagation, as they are key to developing more effective strategies to limit the spread of harmful content on the platform.

To our knowledge, this is the first study to understand the spread of information on WhatsApp at such a scale.  By analyzing the spread of harmful content, this work provides unique insights into how harmful content propagates more broadly and deeply compared to normal messages.

\section{Contributions and Related Work}
Harmful messages on social networks are becoming a major issue, with a significant body of research focused on public social media platforms such as X (formerly Twitter) and Facebook. \citet{Nir2019} find that on Twitter, misinformation exposure is highly concentrated, with just 1\% of users encountering 80\% of false information. Similarly, \citet{Jennifer2024} study vaccine misinformation on Facebook and find that reports implying vaccines are harmful, not being flagged by fact checkers, had a wider reach and significantly reduced people's willingness to get vaccinated. In another study, \citet{Goel2023} highlight the role of hatemongers in spreading hateful speech, noting that these individuals tend to cluster together and form stronger connections within social networks, thus amplifying the spread of harmful content. Likewise, \citet{Hristakieva2022} indicate that propaganda spreads more effectively on social media when coordinated between communities.

Although much of the existing research has focused on public platforms like Twitter and Facebook, WhatsApp has recently emerged as another powerful tool to spread harmful messages. However, research on WhatsApp is limited due to several constraints, such as end-to-end encryption, which makes data collection and analysis particularly challenging. To address these limitations, most studies have focused on public WhatsApp groups. For example, \citet{garimella2018} propose a generalized method for collecting data from the public WhatsApp groups using Selenium scripts to search for publicly available group invite links through Google. Building on this method, \citet{saha2021} conduct the first large-scale analysis of fear speech in public WhatsApp groups discussing politics in India, and find that such messages spread rapidly and are harder to detect due to their low toxicity. Similarly, \citet{resende2019} examine the dissemination of information in political WhatsApp groups in Brazil, focusing on two social events: the national truck drivers strike and the Brazilian presidential campaign. They discover that misinformation, particularly in the form of images, spreads widely between groups and across platforms during these events. However, focusing solely on public groups is insufficient, as it overlooks the private groups where much of WhatsApp's typical usage occurs. In response, this paper constructs a novel dataset that emphasizes private groups, providing a more accurate reflection of how harmful content spreads on WhatsApp.

Accurately identifying harmful messages remains a pressing issue, as conventional machine learning techniques, though effective on public platforms, are often not scalable for detecting harmful content at a large scale on platforms like Twitter or Facebook, and they are not applicable at all on private platforms like WhatsApp due to end-to-end encryption and other privacy barriers. Given these challenges, researchers have increasingly turned their attention to the structural characteristics of message dissemination as an alternative approach to studying harmful content. Analyzing the structural differences between harmful and regular messages has become crucial in this context.

Studies consistently show that harmful messages spread more effectively than regular ones, highlighting clear structural differences. For example, \citet{Soroush2018} find that false news on Twitter spreads faster, deeper and more broadly than true news. Similarly, \citet{Mathew2019} demonstrate that hateful speech travels farther, spreads faster, and reaches a much wider audience than regular messages. Extending these findings, \citet{Maarouf2024} analyze retweet cascades of hateful speech on Twitter and discover that hateful content forms cascades that are 3.5 times larger in size and have 1.2 times greater structural virality, defined as the average distance between all pairs of nodes \cite{Goel2016}, compared to normal content.

In contrast, the structural characteristics of the dissemination of information on WhatsApp are underexplored. To our knowledge, only \citet{caetano2019} analyze attention cascades, which begin when a user introduces a topic in a message serving as a starting point for the cascade. Other users contribute by replying either to the original message or to subsequent replies, forming a chain of interactions. \citet{caetano2019} find that attention cascades involving false information in WhatsApp political groups tend to be deeper, reach more users, and last longer than those in non-political groups. However, attention cascades only account for information dissemination within a single group, overlooking cross-group transmission. This cross-group transmission is often crucial for large-scale viral dissemination, where information spreads from one group to another, amplifying its reach. To address this limitation, this paper focuses on cross-group cascades and compares the structural differences between various types of cascades.

However, cascades are complex dynamic entities, and reconstructing the process of cross-group message dissemination is a significant technical challenge in the study of WhatsApp cascades. Several factors limit the reconstruction of cascades. First, it is impossible for anyone, including WhatsApp itself, to fully track the entire message transmission process. As a result, we rely on sampling methods, where specific groups are chosen, and by detecting when messages appear in these groups, we attempt to infer the structural characteristics of the complete information cascade. We adopt a two-pronged approach: First, we reconstruct the transmission process between the observed and sampled group nodes. Second, after obtaining an incomplete, partially observed cascade, we estimate the structural parameters of the complete cascade. To address the first problem, we use two different methods to construct the \textbf{influence cascade}. We begin by applying the algorithm proposed by \citet{Maunuel2012}, which reconstructs the diffusion process of an incomplete influence cascade by leveraging the observed message timestamps across groups. This method accounts for both external influences and missing nodes, making it particularly well-suited to our setting with sparse observations. Subsequently, we adopt the algorithm introduced by \citet{Eldar2011}, which estimates the expected structural properties of the incomplete cascade—such as the number of nodes and edges—based on a theoretical tree model and known sampling probability (see Section~\ref{methodology} for details). By minimizing the discrepancy between these expected values and the observed data, the algorithm provides accurate estimates of the tree model’s structural parameters, even when up to 90\% of the cascade is unobserved. An alternative line of work models cascade reconstruction as a Steiner tree problem, where the goal is to find a minimum-cost tree that spans all observed nodes while respecting their temporal order \cite{xiao2018reconstructing,rozenshtein2016reconstructing}. However, such methods are less suitable for our application due to their limited capacity to cope with extensive missing data—an inherent characteristic of our dataset, as elaborated in the next section.


\section{Dataset}
\label{sec:dataset}


We collected a dataset of WhatsApp group messages through data donations from 3,500 users in the northern Indian state of Uttar Pradesh, corresponding to 5,953 WhatsApp groups. The data collection spanned from October 2023 to June 2024, yielding more than 5 million messages. To ensure a representative sample of villages and capture diverse demographics, we carefully designed our sampling method.

\noindent\textbf{Sampling Procedure}.
Our sampling procedure involved randomly selecting 10 districts within Uttar Pradesh and then randomly picking 10 villages from each chosen district. For each selected village, we obtained census data to establish the baseline distribution of the population in terms of age, caste, and religion.\footnote{Due to practical limitations during our data donation pilot, we did not attempt to obtain a sample representative of gender.}
Based on this distribution, our survey team visited each village and sought the consent of the participants until we filled the quotas corresponding to the age, caste, and religion demographics. We opted for a quota sampling method instead of a purely random sample due to practical considerations related to the uptake of our data donation process.

\noindent\textbf{Data Donation Process}.
The on-the-ground protocol involved surveyors reaching out to participants to obtain informed consent and explain our data collection and anonymization protocols. We used a custom-built data donation tool to facilitate users in donating their WhatsApp group data. Only groups with more than five participants and a certain level of activity were eligible for donation.
Upon completion of the donation process, our tool collected the following data: (i) All messages from the two months preceding and the two months following the date of donation. (ii) Anonymized contacts from the user's phone contacts; and, (iii) Anonymized group membership information.

\noindent\textbf{Dataset Statistics}. The collected dataset comprises over 5 million messages from more than 5,900 WhatsApp groups during October 2023 to June 2024. The median group size was 104 members, indicating that most groups were large and engaged in discussions around political and religious identity, caste, region, and related topics. To better illustrate the structure of the dataset, Table~\ref{tab:summary_statistics} presents descriptive statistics for the dataset. Specifically, we report the total number of messages and the total number of groups involved, disaggregated by modality, content type, and forwarding score. These statistics help characterize the prevalence and potential reach of different types of messages, including how many were forwarded multiple times, with a particular emphasis on those considered viral (i.e., forwarded five or more times).

\begin{table}[htbp]
\centering
\caption{Summary statistics of messages by modality, content type, and forwarding score.}
\label{tab:summary_statistics}
\resizebox{\linewidth}{!}{
\begin{tabular}{llrr}
\toprule
\textbf{Category} & \textbf{Item} & \textbf{Messages} & \textbf{Groups} \\
\midrule
\textbf{Modality} & Text & 2,650,441 & 5,631 \\
& Image & 2,100,004 & 5,539 \\
& Video & 404,434 & 4,165 \\
\midrule
\textbf{Content Type} & Misinformation & 401 & 122 \\
& Hate & 111 & 49 \\
& Propaganda & 116 & 59 \\
\midrule
\textbf{Forwarding Score} & 0 (Not forwarded) & 3,629,277 & 5,953 \\
& 1 & 803,607 & 4,295 \\
& 2 & 179,204 & 3,293 \\
& 3 & 70,262 & 2,578 \\
& 4 & 34,192 & 2,046 \\
& $\geq$5 (Viral) & 74,037 & 2,364 \\
\bottomrule
\end{tabular}
}
\end{table}

\noindent\textbf{Annotations}. We annotated a subset of messages to identify content that contains misinformation, hate speech, and political propaganda. Specifically, we annotated all 2,019 pieces of content that were marked as ``forwarded many times" and shared during October to December 2023.\footnote{We used definitions of misinformation, hate speech, and propaganda from the Facebook Community Standards documents (\url{https://transparency.meta.com/policies/community-standards/misinformation}).} The annotations were performed by a professional fact-checker who was well-versed with the content and cultural context of the data. It is important to note that, due to privacy restrictions, we were unable to classify all messages in the dataset as harmful or not. Therefore, we focused our analysis on messages that had been forwarded many times---specifically, those with a forwarding score of five or more---and applied our harmful message classification to this subset. Within this smaller but highly relevant subset of viral messages, we manually annotated 1,149 instances of \textit{viral normal} messages, which primarily consisted of advertising, entertainment, greetings (e.g., ``good morning''), informational, inspirational, local/national news, and religious content. In addition, we labeled 401 instances of misinformation, 111 instances of hate speech, and 116 instances of propaganda. This careful annotation process ensures that our conclusions are not biased by the selection of highly forwarded messages. Furthermore, for the remaining unlabeled data, we also computed the structural parameters \(b\) and \(h\). Since misinformation typically constitutes a small proportion of all messages, the structural properties of these unlabeled cascades are expected to resemble those of \textit{non-viral normal} messages (i.e., those not forwarded many times). This provides further support for the robustness of our findings.

Out of the annotated messages, we identified 401 instances of misinformation, 111 instances of hate speech, and 116 instances of propaganda.\footnote{This sample of misinformation, hate and propaganda is not an ideal sample since it does not cover all the time period of our messages. However, an annotation of thousands of pieces of content is very tedious and time consuming, requires experts, and does not scale well. We are in the process of collecting expert annotations for the rest of our dataset and will update the manuscript with numbers that compare misinformation on the entire dataset.}

\noindent\textbf{Ethical Considerations}.
The data collection process was approved by the Institutional Review Board (IRB) at multiple participating institutions (details anonymized). We took extreme care to minimize the amount of data collected and to protect personal information. Personal identifiers, including phone numbers, emails, and faces in images, were anonymized before storage on our servers.
Importantly, the data analyzed in this study did not include the content of the messages; only metadata were analyzed to ensure the privacy of participants.

\section{Methodology}\label{methodology}
In this paper, we focus on the structural characteristics of information dissemination, particularly the breadth and depth of cascades. We assume that the information propagates in a tree-like structure and attempt to reconstruct the dissemination process based on the available data. A key challenge lies in the fact that our dataset does not allow us to fully trace the entire information transmission process. This limitation is almost inevitable when analyzing platforms such as WhatsApp, where even the platform itself is unable to track the entire path of information dissemination. Even if we assume that the platform can trace all instances of the same content, the information could originate from two distinct transmission paths, with each source independently influenced by external factors. As such, accurately reconstructing the entire cascade of information transmission is almost impossible. Instead, we focus on providing reliable estimates for certain structural characteristics of the cascades with missing data.

Another challenge is that in our dataset, we can only observe when each group receives a message, but not the actual transmission paths between them. Ideally, a cascade $c$ would consist of triples of the form $(u, v, t_v)_c$, indicating that a message appeared in group $u$ and was subsequently forwarded to group $v$, which received it at time $t_v$. However, such detailed propagation information is not available; we only observe pairs $(v, t_v)_c$, meaning that group $v$ received the message at time $t_v$, without knowing from which group it was transmitted. To address this limitation, we adopt the method of \citet{Maunuel2012} to reconstruct the missing transmission paths from the observed timing data. The core idea is to estimate the probability that a directed edge exists between two groups based on the temporal proximity of their adoption times. This allows us to reconstruct a partial \textit{influence cascade}, which we use to infer the underlying structure of the complete diffusion process.

\subsection{Influence Cascade Reconstruction}

The \textit{influence cascade} model aims to reconstruct the underlying diffusion process based on when messages are observed in different groups. In our setting, we observe the times at which messages first appear in different groups, but the specific diffusion paths between groups are not observable. To model this, we follow the framework proposed by \citet{Maunuel2012}, which assumes that information diffuses over an unknown, static, directed network, where each group can potentially influence others. Each observed cascade is the result of a hidden diffusion process over this network.

We assume that each message propagates over the network as a tree-structured cascade, where each infected group is influenced by exactly one predecessor. The likelihood of a group $v$ being influenced by group $u$ depends on their temporal proximity, modeled via an exponential decay function:

\[
P_c(u, v) \propto e^{-\Delta_{u,v}/\alpha}
\]
where $\Delta_{u,v} = t_v - t_u$ is the time delay and $\alpha$ controls the rate of decay.

Given this transmission model, the likelihood that a cascade $c$ propagates through a specific tree $T = (V_T, E_T)$ can be written as:

\[
P(c \mid T) = \prod_{(u,v) \in E_T} \beta P_c(u, v) \prod_{\substack{u \in V_T,\\ (u,x) \in E \setminus E_T}} (1 - \beta)
\]

Here, $\beta$ is the probability that an infected node successfully transmits the message to a neighbor, while $(1 - \beta)$ penalizes non-transmissions to reachable neighbors. In the original algorithm of \citet{Maunuel2012}, the possibility of missing nodes is also considered, acknowledging that the observed cascade may only reflect part of the true diffusion process. To address this, the computation of \( P(c \mid T) \) is extended to incorporate external influence nodes, enabling the model to account for incomplete cascades. For further details, we refer the reader to Section~3.1 of \citet{Maunuel2012}.

Since we do not observe the actual diffusion paths, we marginalize over all possible trees consistent with the observed infection times to compute the cascade likelihood:

\[
P(c \mid G) = \sum_{T \in \mathcal{T}_c(G)} P(c \mid T) P(T \mid G),
\] where \(\mathcal{T}_c(G)\) is the set of all directed trees in \(G\) that are consistent with the infection times observed in cascade \(c\).

Assuming that the cascades are conditionally independent given the network $G$, the likelihood of observing the entire cascade collection $C$ is:
\[
P(C \mid G) = \prod_{c \in C} P(c \mid G).
\]
Finally, our objective is to infer the underlying diffusion network $\hat{G}$ and the cascade trees $\hat{T}$ by maximizing this likelihood. To ensure the validity of the cascades, we preprocess the data by keeping only the earliest timestamp when a message appears in a group. Messages that appear in only a single group are excluded, as they do not contribute to the inference of inter-group diffusion patterns. This process allows us to reconstruct incomplete influence cascades that reflect meaningful information propagation across multiple groups.

Although this reconstruction is based on incomplete data, it still allows us to examine key structural properties such as \textbf{maximum breadth} and \textbf{depth}. These properties differ from the parameters $b$ and $h$ used in the tree model, which are also referred to as breadth and depth, but represent different aspects of the cascade. Specifically, depth in this context is defined as the total number of edges from the root node to the leaf node along the longest path in the incomplete influence cascade, while breadth refers to the number of messages at a particular depth level in the cascade. Our focus is on the maximum breadth, defined as the largest number of messages at any depth level.  It is worth noting that the terms depth and breadth used here differ slightly from the depth and breadth described in the tree model later in the paper, and care should be taken to distinguish between these two contexts to avoid potential confusion.

\begin{figure}[H]
  \centering 
  \begin{subfigure}[t]{0.48\linewidth}
    \centering
    \includegraphics[width=\linewidth, keepaspectratio]{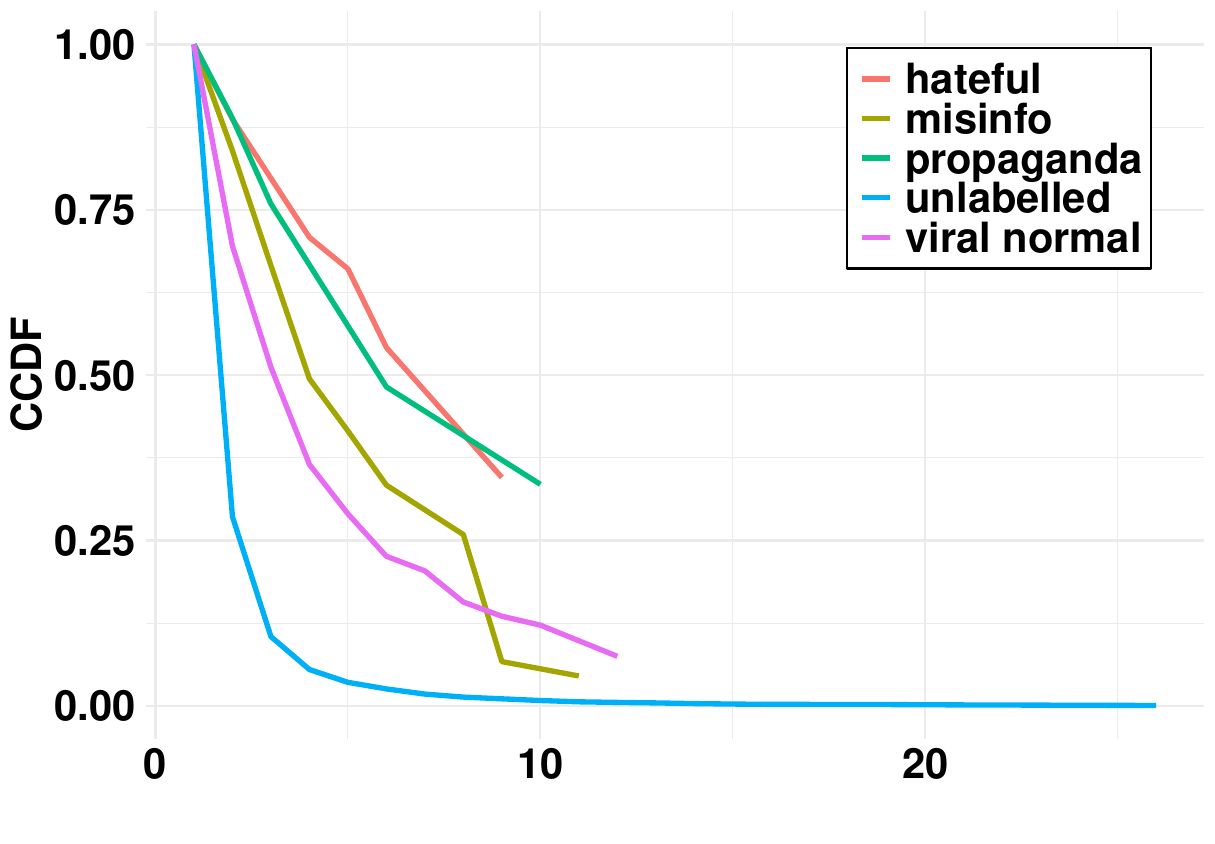}
    \caption{Breadth}
    \label{fig:ccdf_breadth}
  \end{subfigure}
  \hfill
  \begin{subfigure}[t]{0.48\linewidth}
    \centering
    \includegraphics[width=\linewidth, keepaspectratio]{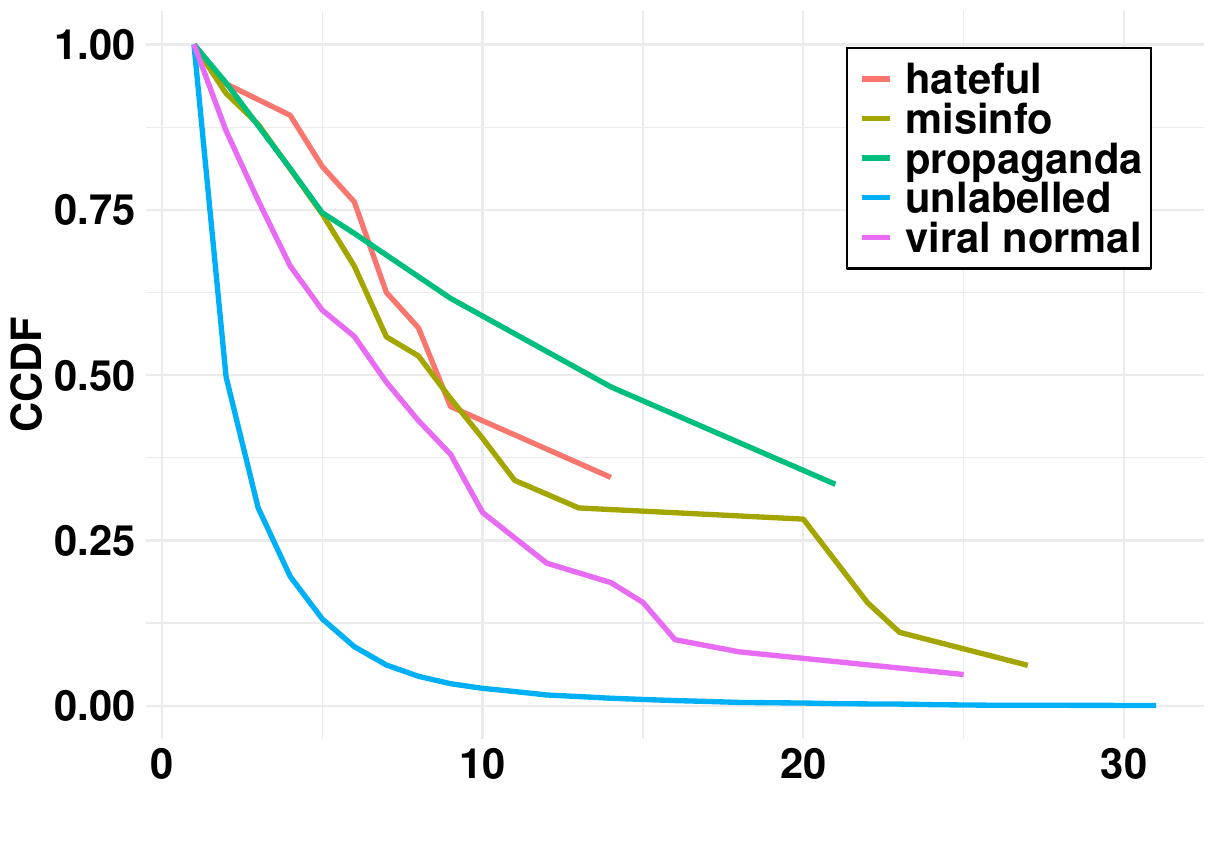}
    \caption{Depth}
    \label{fig:ccdf_depth}
  \end{subfigure}
  \caption{Comparison of CCDF for Breadth and Depth}
\end{figure}

By analyzing the maximum breadth and depth for each message type, we aim to demonstrate that even in the sampled dataset, harmful messages, viral normal messages, and unlabeled messages exhibit significant differences in their propagation patterns. To support this, we present complementary cumulative distribution function (CCDF) plots to compare how different message types spread across groups. As shown in Figures \ref{fig:ccdf_breadth} and \ref{fig:ccdf_depth}, harmful messages demonstrate significantly greater breadth and depth compared to viral normal messages and unlabeled messages, indicating their broader and deeper propagation between groups. In addition, propaganda messages exhibit greater breadth and depth compared to misinformation. These findings highlight that, despite the limitations of the sampled data, the observed structural differences are robust and consistent.

\subsection{Inferring Structural Parameters of the Complete Cascade}
Using this partially observed cascade dataset, our goal is to infer the structural properties of the complete cascade tree. Specifically, given only a subset \(C'\) of the complete cascade \(C\), we aim to estimate key parameters of the underlying diffusion process, such as its breadth (\(b\)) and depth (\(h\)).

To accomplish this, we adopt the model-based inference method proposed by \citet{Eldar2011}, which assumes that the complete cascade follows a tree structure and that each node is observed independently with probability \(p\). We denote the sampled tree as \(\Gamma(p, b, h)\), where \(b\) is the breadth and \(h\) is the depth of the tree.

\begin{figure}[h]
    \centering
    \includegraphics[width=0.4\textwidth]{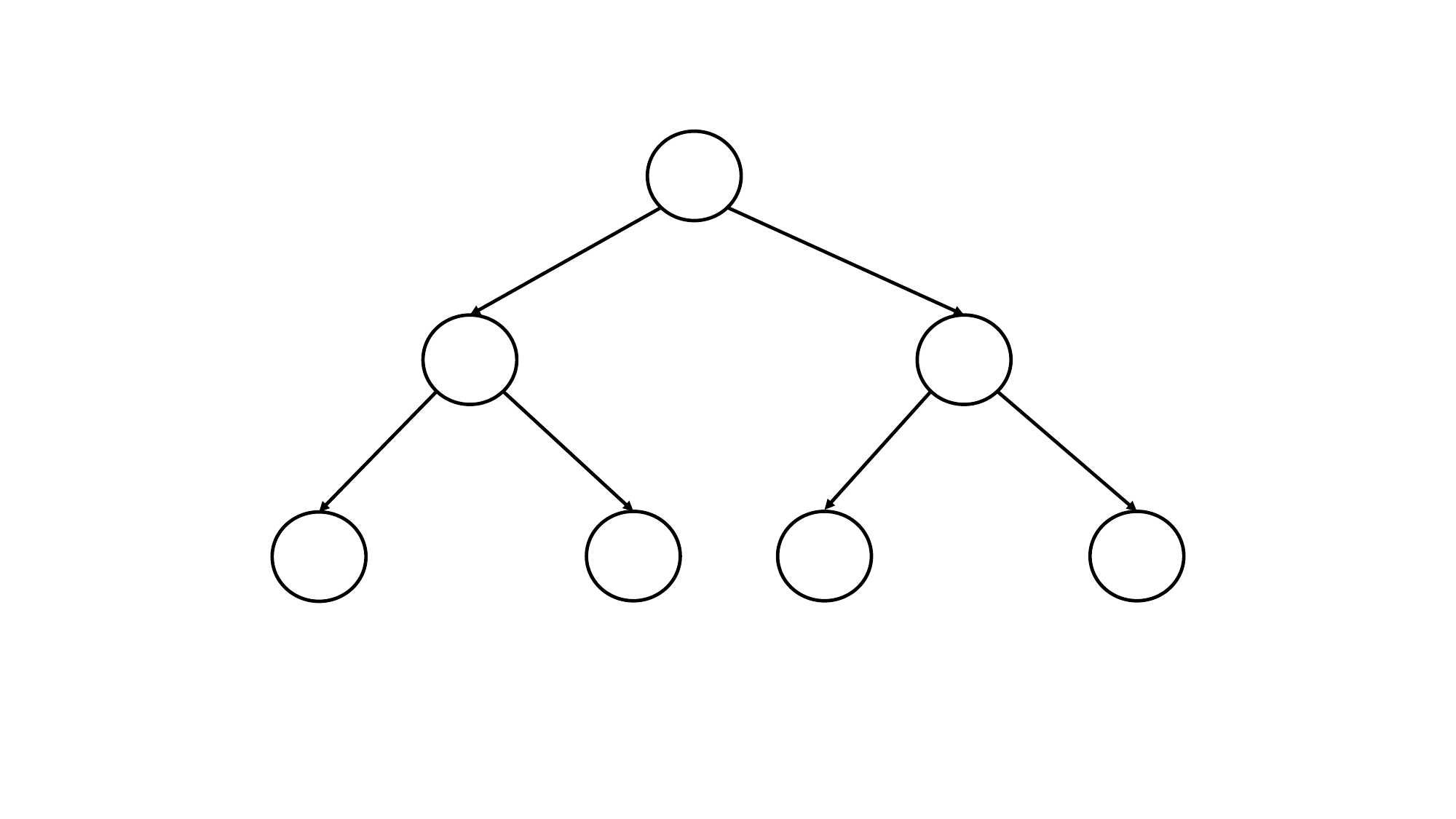}
    \caption{A tree cascade with breadth $b=2$ and depth $h=2$.}
    \label{fig:tree_model}
\end{figure}

Figure~\ref{fig:tree_model} provides a visual illustration of this tree model, where \(b = 2\) represents the number of children each internal node produces, and \(h = 2\) denotes the number of layers from the root to the leaves. This structure models a typical influence cascade process with hierarchical spread.

The key idea is to compute theoretical expectations for several structural properties of the sampled tree \(\Gamma(p, b, h)\), and to estimate \(b\) and \(h\) by minimizing the discrepancy between these theoretical values and the corresponding empirical values measured from \(C'\). This approach enables robust estimation even when a large portion of the cascade is missing. Details of the matched properties and the derivation of the expectations can be found in \cite{Eldar2011}.

In our case, the parameter \( p \) is the sampling probability of each group, which we can calculate based on the data collection process (as described in Section~3). Practically, we set the value of \( p \) to 0.02 (2\%) given that we were sampling roughly 1\% of a village and most of the users provided almost 90\% of the groups they had. Since group memberships may overlap across users, the effective group coverage can exceed the nominal sampling rate of 1\%, and we therefore conservatively set \( p = 0.02 \) to reflect this amplification. Then, we can estimate the breadth and depth by minimizing the sum of errors between the values of these properties from our dataset and those calculated from the theoretical sampled tree model \(\Gamma(p, b, h)\).
Note that the exact value of \( p \) is context dependent --- i.e., might vary from village to village and is a rough parameter which we experimented with. We provide experiments varying the value of \( p \) in Appendix B. We can see in Section Appendix B that our results are robust to the value of \( p \).

\subsection{Population-Level Impact Estimation}
Based on the estimates of the influence cascade, we can derive the parameters required for the complete tree model. This allows us to understand the structural characteristics of specific types of information transmission, such as whether misinformation has greater breadth or depth compared to normal messages. Additionally, we can calculate their respective reach, such as how many groups an average misinformation message is expected to propagate through. By analyzing our dataset, we can also estimate the distribution of group sizes. This enables us to combine the number of groups traversed by a complete information cascade with the size of those groups, providing insights into the population-level impact of specific information types, namely, how many people are ultimately affected by different types of message.

\section{Validation via Synthetic Dataset}
To evaluate the accuracy of the \((b, h)\) estimation procedure, we conducted a simulation study using synthetic cascades. We first generated an Erdős--Rényi network to represent the underlying contact structure. Each edge was assigned a random transmission delay drawn from an exponential distribution, and the probability of information transmission between two nodes was modeled as an exponentially decreasing function of this delay.

Based on this transmission model, we simulated 100 influence cascades for two types of messages, distinguished by their maximum allowable cascade durations. From each complete cascade, we sampled nodes at rates ranging from 2\% to 5\%, retaining only their activation timestamps to construct \textit{partial influence cascades}.

We then applied the cascade reconstruction algorithm of Gomez-Rodriguez et al.\ to infer the likely diffusion structure from the sampled data. Using the reconstructed cascades, we estimated the structural parameters \((b, h)\) by minimizing the error between the observed statistics and those predicted by the theoretical sampling model \(\Gamma(p, b, h)\). The estimated values were then compared to the ground-truth parameters used in the cascade generation process.

\begin{table}[ht]
\centering
\caption{Validation of $(b, h)$ Estimation on Simulated Cascades}
\resizebox{\columnwidth}{!}{
\begin{tabular}{llcccccc}
\toprule
\textbf{Cascade Type} & \textbf{Sampling} & $b$ & $\hat{b}$ & \textbf{Rel. Err.} & $h$ & $\hat{h}$ & \textbf{Rel. Err.} \\
\midrule
Long-duration & 0.02 & 1.962 & 2.035 & 0.037 & 10.477 & 10.243 & 0.022 \\
Long-duration & 0.03 & 1.960 & 2.068 & 0.055 & 10.443 & 10.109 & 0.032 \\
Long-duration & 0.04 & 1.960 & 2.010 & 0.026 & 10.443 & 10.321 & 0.012 \\
Long-duration & 0.05 & 1.960 & 2.003 & 0.022 & 10.443 & 10.431 & 0.001 \\
\midrule
Short-duration & 0.02 & 1.933 & 1.750 & 0.095 & 6.240 & 7.681 & 0.231 \\
Short-duration & 0.03 & 1.931 & 1.765 & 0.086 & 6.087 & 7.138 & 0.173 \\
Short-duration & 0.04 & 1.910 & 1.712 & 0.104 & 5.977 & 7.203 & 0.205 \\
Short-duration & 0.05 & 1.918 & 1.677 & 0.126 & 5.792 & 7.253 & 0.252 \\
\bottomrule
\end{tabular}
}
\label{tab:validation}
\end{table}

To assess the reliability of our cascade reconstruction and \((b, h)\) estimation methods, we conducted a validation experiment using synthetic cascades generated on an Erdős--Rényi network. The results, summarized in Table~\ref{tab:validation}, report estimation accuracy for two types of cascades with differing structural properties, distinguished by their maximum cascade durations.

For cascades with longer durations and greater depth (with ground truth parameters approximately \(b \approx 1.96\), \(h \approx 10.44\)), the estimated values \(\hat{b}\) and \(\hat{h}\) closely matched the true parameters across all tested sampling rates (2\% to 5\%). The relative errors were consistently low, particularly for depth, which remained below 3.2\% and dropped to just 0.1\% at the 5\% sampling rate. These results indicate that the proposed method is highly accurate in reconstructing deeper and more temporally dispersed cascades.

For shorter-duration cascades (with ground truth \(b \approx 1.92\), \(h \approx 6.0\)), estimation accuracy was slightly lower. The estimated \(\hat{b}\) values remained relatively close to the true branching factor, with relative errors between 8.6\% and 12.6\%. However, the depth estimates \(\hat{h}\) showed greater variability and tended to overestimate the true depth, resulting in higher relative errors ranging from 17.3\% to 25.2\%. This increased variance is likely attributable to the inherently shallower structure of these cascades, where small reconstruction errors can lead to proportionally larger deviations.

Overall, these validation results support the robustness of our reconstruction pipeline. While estimating depth is more challenging for shorter cascades, the method demonstrates strong performance across conditions and is particularly effective in capturing the structure of longer, more complex cascades.

\section{Results}
In this analysis, our objective is to understand the dissemination patterns of different message types, including harmful messages such as misinformation, hate speech, and propaganda, as well as normal messages. We start by analyzing the structural characteristics of message dissemination through the influence cascade. In addition, we examine how both message modality—including chat, image, and video—and content type—namely misinformation, hateful speech, propaganda, and normal messages—affect dissemination patterns. We also incorporate the forwarding score as a proxy for message-level virality to better quantify the extent of spread. Finally, we provide population-level estimates to quantify the total number of people typically affected by each message type, combining group size data with the average number of groups through which a message passes. This comprehensive approach allows us to identify structural differences in the spread of harmful content, offering key insights into their broader societal impact.

\subsection{Influence Cascade}

Based on insights into structural characteristic differences from incomplete influence cascades, we can extend the analysis to examine the structural characteristics of complete influence cascades across five message types: misinformation, hate speech, propaganda, viral normal and unlabelled messages. Specifically, we model the complete influence cascade using a tree structure, characterized by two key parameters: the branching factor, \(b\) (breadth), which indicates the average number of connections a node generates, and the depth, \(h\), which represents the maximum number of layers in the tree, or how deep the information travels. We assume that the incomplete influence cascade observed in our dataset is a sampled version of the complete cascade, with some connections or nodes missing due to the limitations of the data. 

\begin{table}[h]
  \centering
  \caption{Mean and Standard Deviation of Parameters by Content Type}
  \label{tab:influence_cascade}
  \begin{tabular}{lcccc}
    \toprule
  Content Type & $\mu_b$ & $\sigma_b$ & $\mu_h$ & $\sigma_h$ \\
    \midrule
    Hateful     & 3.78 & 0.575 & 4.89 & 0.244 \\
    Misinformation & 3.68 & 0.579 & 4.86 & 0.261 \\
    Propaganda   & 3.82 & 0.660 & 4.92 & 0.298 \\
    Viral Normal      & 3.47 & 0.612 & 4.77 & 0.273 \\
    Unlabeled      & 2.85 & 0.388 & 4.50 & 0.167 \\ \bottomrule
  \end{tabular}
\end{table}

For each type of message, we estimate the values of \(b\) and \(h\) to capture the spread dynamics within the influence cascade. The results, shown in Table~\ref{tab:influence_cascade}, indicate that harmful messages, including misinformation, hate speech, and propaganda, exhibit higher values of \(b\) and \(h\) compared to normal messages. This suggests that harmful content spreads more broadly and deeply, reaching a wider audience and traveling further into the network.

Specifically, while all harmful messages show higher breadth and depth compared to normal messages and normal messages with a forwarding score greater than or equal to five, there are notable differences among the harmful message types. Propaganda has the highest values for both breadth (\(b = 3.82\)) and depth (\(h = 4.92\)), followed closely by hate speech with a breadth of \(b = 3.78\) and depth of \(h = 4.89\). Misinformation, though still significantly larger than normal viral messages, has slightly lower values (\(b = 3.68\) and \(h = 4.86\)). This suggests that, although misinformation spreads widely, propaganda and hate speech tend to spread even more broadly and deeply on WhatsApp.

\subsection{Wilcoxon Rank-Sum Test Results for the Breadth and Depth of Influence Cascades}
To ensure the robustness of our results, we performed one-sided Wilcoxon rank sum tests to compare the breadth and depth of cascades across different types of content for influence cascades, such as testing hypotheses whether the depth of misinformation is significantly greater than that of normal messages, as shown in Tables \ref{tab:wilcox_rank_sum_test_results_breadth} and \ref{tab:wilcox_rank_sum_test_results_depth}. The results show that, for both types of influence cascades, there are significant structural differences between harmful content types (including misinformation, hate speech, and propaganda) and viral normal messages or unlabeled messages.

For the breadth of cascades, the comparisons between misinformation, hate speech, propaganda, and viral normal messages consistently show highly significant differences (p-value $<$ 1e-5). Furthermore, pairwise comparisons between propaganda and misinformation reveal notable differences, indicating that even among harmful messages, the structural properties of dissemination vary significantly.

For the depth of cascades, similar patterns emerge. Harmful messages such as misinformation, hate speech, and propaganda exhibit significantly deeper cascades compared to viral normal messages, with p-values $<$ 1e-5 in most comparisons. This indicates that harmful messages penetrate the network more deeply than regular content. In addition, the difference between propaganda and misinformation remains significant, further emphasizing the unique structural dynamics between different types of harmful content.

These results reinforce the conclusion that harmful messages spread not only more broadly, but also more deeply through networks compared to normal messages. Furthermore, the differences between various types of harmful content suggest that certain types of harmful content, such as propaganda, may be more effective at reaching a broader and deeper audience than others, such as misinformation. The significance of these results across both influence and network cascades ensures the robustness of our findings, confirming the reliability of our cascade reconstruction methods in capturing the structural differences in message dissemination.

\begin{table}[h]
  \centering
  \caption{Wilcoxon Rank-Sum Test Results for the Breadth of Influence Cascades}
  \label{tab:wilcox_rank_sum_test_results_breadth}
  \begin{tabular}{lcc}
    \toprule
    Content Type Comparison & Test Statistic & p-value \\
    \midrule
    Misinfo - Unlabeled & 728576807.00 & $<1\mathrm{e}{-5}$ \\
    Hateful - Unlabeled & 100453626.00 & $<1\mathrm{e}{-5}$ \\
    Propa - Unlabeled & 132289961.00 & $<1\mathrm{e}{-5}$ \\
    Misinfo - Viral Normal  & 1209173.00 & $<1\mathrm{e}{-5}$ \\
    Hateful - Viral Normal  & 173794.50 & $<1\mathrm{e}{-5}$ \\
    Propa - Viral Normal  & 236950.50 & $<1\mathrm{e}{-5}$ \\
    Hateful - Misinfo & 114896.50 & 0.0215 \\
    Propa - Misinfo & 158536.00 & 0.0006 \\
    Propa - Hateful & 20567.50 & 0.0563 \\
    \bottomrule
  \end{tabular}
\end{table}

\begin{table}[h]
  \centering
  \caption{Wilcoxon Rank-Sum Test Results for the Depth of Influence Cascades}
  \label{tab:wilcox_rank_sum_test_results_depth}
  \begin{tabular}{lcc}
    \toprule
    Content Type Comparison & Test Statistic & p-value \\
    \midrule
    Misinfo - Unlabeled & 728320653.00 & $< 1\mathrm{e}{-5}$ \\
    Hateful - Unlabeled & 100515405.00 & $< 1\mathrm{e}{-5}$ \\
    Propaganda - Unlabeled & 131765052.00 & $< 1\mathrm{e}{-5}$ \\
    Misinfo - Viral Normal  & 1206402.00 & $< 1\mathrm{e}{-5}$ \\
    Hateful - Viral Normal  & 173893.50 & $< 1\mathrm{e}{-5}$ \\
    Propaganda - Viral Normal & 235748.50 & $< 1\mathrm{e}{-5}$ \\
    Hateful - Misinfo & 107938.50 & 0.2661 \\
    Propaganda - Misinfo & 159914.00 & 0.0002 \\
    Propaganda - Hateful & 20582.50 & 0.0548 \\
    \bottomrule
  \end{tabular}
\end{table}

\subsection{The Impact of Message Modality on Dissemination Patterns}
\begin{figure}[h]
  \centering
  \includegraphics[width=\linewidth]{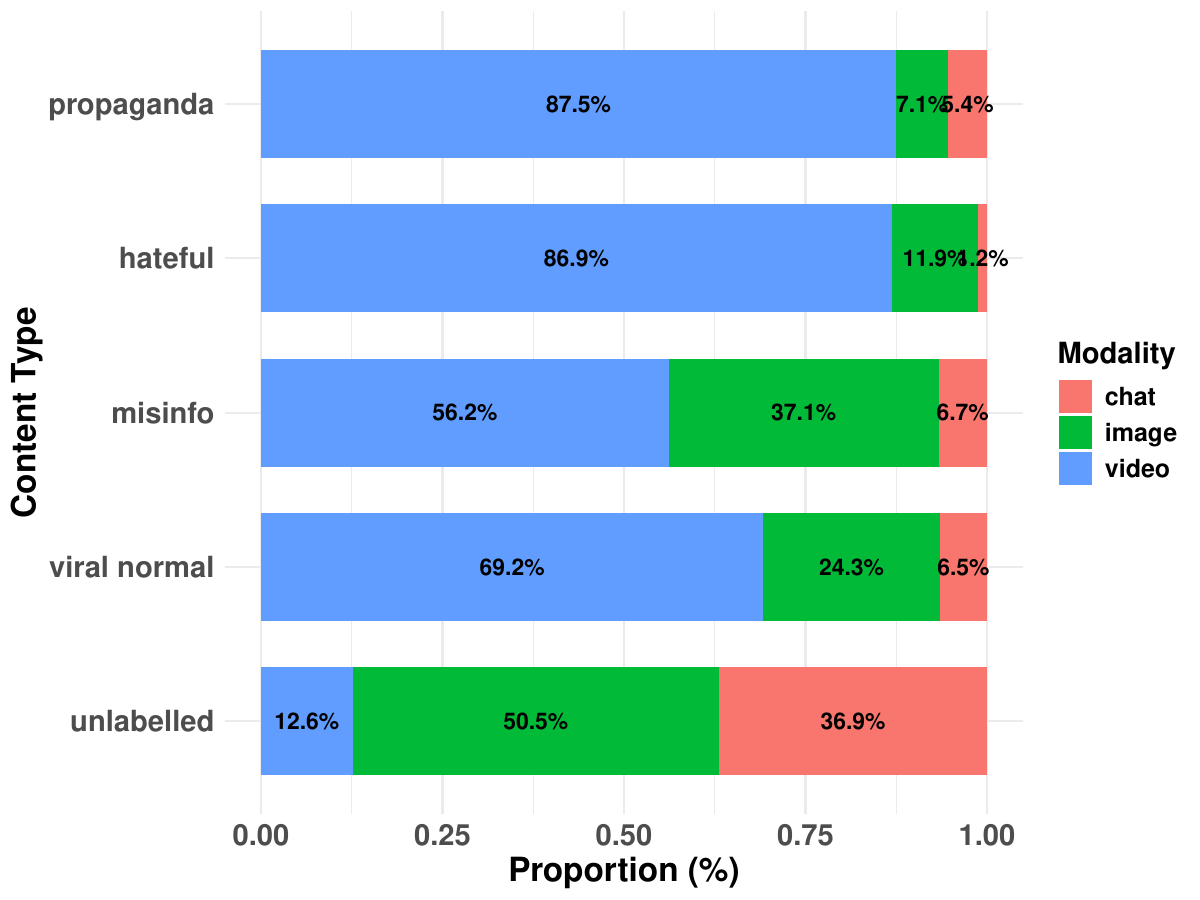}
  \caption{Proportion of different modality types within each content type.}
  \label{fig:modality_content_type_plot}
\end{figure}

In this section, we explore the reasons behind the broader dissemination of harmful messages—such as hateful content, misinformation, and propaganda—compared to other normal and unlabeled viral messages. Based on our analysis of the dataset, Figure \ref{fig:modality_content_type_plot} highlights significant variations in the distribution of dissemination modalities (chat, image, video) between different types of content. For example, video is the dominant modality for both hateful speech and political propaganda, comprising 86.9\% and 87.5\% of their respective message distributions. In contrast, normal viral content and viral misinformation exhibit a more balanced distribution, with a larger portion consisting of images. These differences suggest that hateful speech and political propaganda rely heavily on video dissemination, while misinformation is more evenly split between image and video, with a smaller emphasis on video. This distribution pattern suggests that the modality through which messages are shared may influence the structure of their transmission, potentially contributing to the wider reach observed in harmful messages.

To test this hypothesis, we classified messages by modality (chat, image, video) and analyzed the breadth and depth of influence cascades for each modality, as shown in Table \ref{tab:influence_cascade_modality}. Our results indicate that video messages consistently exhibit a greater breadth of dissemination compared to chat and image messages in both cascade models. However, despite these differences, the structural characteristics (breadth, depth, and complexity) between chat, video, and image do not vary significantly. This suggests that while modality may contribute to some differences in dissemination, it alone is not sufficient to explain why harmful messages, such as propaganda and hateful speech, spread more widely and deeply than normal messages. 


\begin{table}[h]
  \centering
  \caption{Comparison of Influence Cascade for different modalities}
  \label{tab:influence_cascade_modality}
  \begin{tabular}{lcccc}
    \toprule
  Content Type & $\mu_b$ & $\sigma_b$ & $\mu_h$ & $\sigma_h$ \\
    \midrule
    chat  & 2.84 & 0.389 & 4.49 & 0.167 \\
    video & 3.06 & 0.524 & 4.59 & 0.225 \\
    image  & 2.81 & 0.332 & 4.48 & 0.142\\
    \bottomrule
  \end{tabular}
\end{table}
Therefore, the broader dissemination of harmful messages likely results from a combination of factors beyond just modality, indicating the need for further research to explore deeper drivers of dissemination dynamics in harmful content.

To better understand the drivers of cascade virality, we conducted regression analyses examining the effects of both \textit{modality} (i.e., \emph{video}, \emph{image}, \emph{chat}) and \textit{content type} (i.e., \emph{viral normal}, \emph{propaganda}, \emph{hateful}, \emph{misinformation}) on two key virality metrics: breadth ($b$), defined as the number of unique groups reached and depth ($h$), defined as the maximum time step reached in the inferred cascade.

We estimated the following two linear models:
\begin{align*}
b_i =& \beta_0 + \beta_1 \cdot \mbox{forwarding score}_i + \beta_2 \cdot \mbox{modality}_i \\ &+ \beta_3 \cdot \mbox{content type}_i + \varepsilon_i,
\end{align*}
\begin{align*}
h_i = & \beta_0 + \beta_1 \cdot \mbox{forwarding score}_i + \beta_2 \cdot \mbox{modality}_i \\ &+ \beta_3 \cdot \mbox{content type}_i + \varepsilon_i,
\end{align*} where \emph{forwarding score} was treated as a categorical variable with levels 0, 1, 2, 3, 4, and $5+$. We limited modality to the three most prevalent types and set \emph{normal} as the reference level for content type to highlight differences in virality across message types.

This modeling approach enables us to isolate and quantify the contributions of content semantics and message format to virality, while controlling for the intensity of forwarding behavior. Table~\ref{tab:regression_results_star_se} reports the regression results with standard errors and significance levels.

\begin{table}[htbp]
\centering
\caption{Regression estimates for breadth ($b$) and depth ($h$). Significance levels: *** $p<0.01$, ** $p<0.05$, * $p<0.1$.}
\label{tab:regression_results_star_se}
\resizebox{\linewidth}{!}{
\begin{tabular}{lrrrr}
\toprule
\textbf{Variable} & $b$  & SE($b$) & $h$ & SE($h$) \\
\midrule
(Intercept)               & 2.6882***  & 0.0008 & 4.4282***  & 0.0003 \\
Forwarding Score = 1      & -0.0170*** & 0.0011 & -0.0072*** & 0.0005 \\
Forwarding Score = 2      &  0.0082*** & 0.0020 &  0.0035*** & 0.0008 \\
Forwarding Score = 3      &  0.0254*** & 0.0033 &  0.0108*** & 0.0014 \\
Forwarding Score = 4      &  0.0562*** & 0.0046 &  0.0240*** & 0.0020 \\
Forwarding Score = 5+     &  0.1588*** & 0.0031 &  0.0677*** & 0.0013 \\
Modality = image          & -0.0040*** & 0.0010 & -0.0018*** & 0.0004 \\
Modality = video          &  0.0445*** & 0.0017 &  0.0189*** & 0.0007 \\
Content Type = hateful    &  0.3817*** & 0.0553 &  0.1633*** & 0.0237 \\
Content Type = misinfo    &  0.3251*** & 0.0181 &  0.1409*** & 0.0081 \\
Content Type = propaganda &  0.2932*** & 0.0213 &  0.1264*** & 0.0197 \\
\bottomrule
\end{tabular}
}
\end{table}

The regression results provide strong empirical evidence that content type is the dominant factor in determining cascade virality, even after accounting for modality and forwarding score. Messages labeled as \emph{hateful speech} are associated with an average increase of 0.38 in breadth and 0.16 in depth, both statistically significant at the $p<0.01$ level. Similarly, \emph{misinformation} and \emph{propaganda} cascades also exhibit significantly greater spread: for example, \emph{misinformation} increases breadth by 0.33 and depth by 0.14.

In contrast, modality plays a more modest role. While \emph{video} messages are associated with a small positive increase in both breadth (0.045) and depth (0.019), these effects are considerably smaller than those associated with content semantics. Notably, the \emph{image} modality is negatively associated with both virality measures.

Forwarding score, as expected, is positively and strongly associated with virality. Cascades with a forwarding score of 5 or more show the largest increases in both breadth (0.16) and depth (0.068), confirming the importance of forwarding intensity as a proxy for exposure.

Overall, these findings suggest that although modality contributes to virality, its impact is significantly smaller than that of content type. The semantic nature of the message—whether it conveys misinformation, hate, or propaganda—is the primary determinant of how far and wide a cascade spreads.

\subsection{Population-Level Estimation}
In this final section, we estimate the population-level impact of different message types by approximating how many individuals are typically exposed. Our dataset provides the number of participants in each WhatsApp group, as well as the exact groups through which each message type was propagated. This allows us to obtain the empirical distribution of group sizes for each message type.

\begin{figure}[ht]
    \centering
    \includegraphics[width=0.5\textwidth]{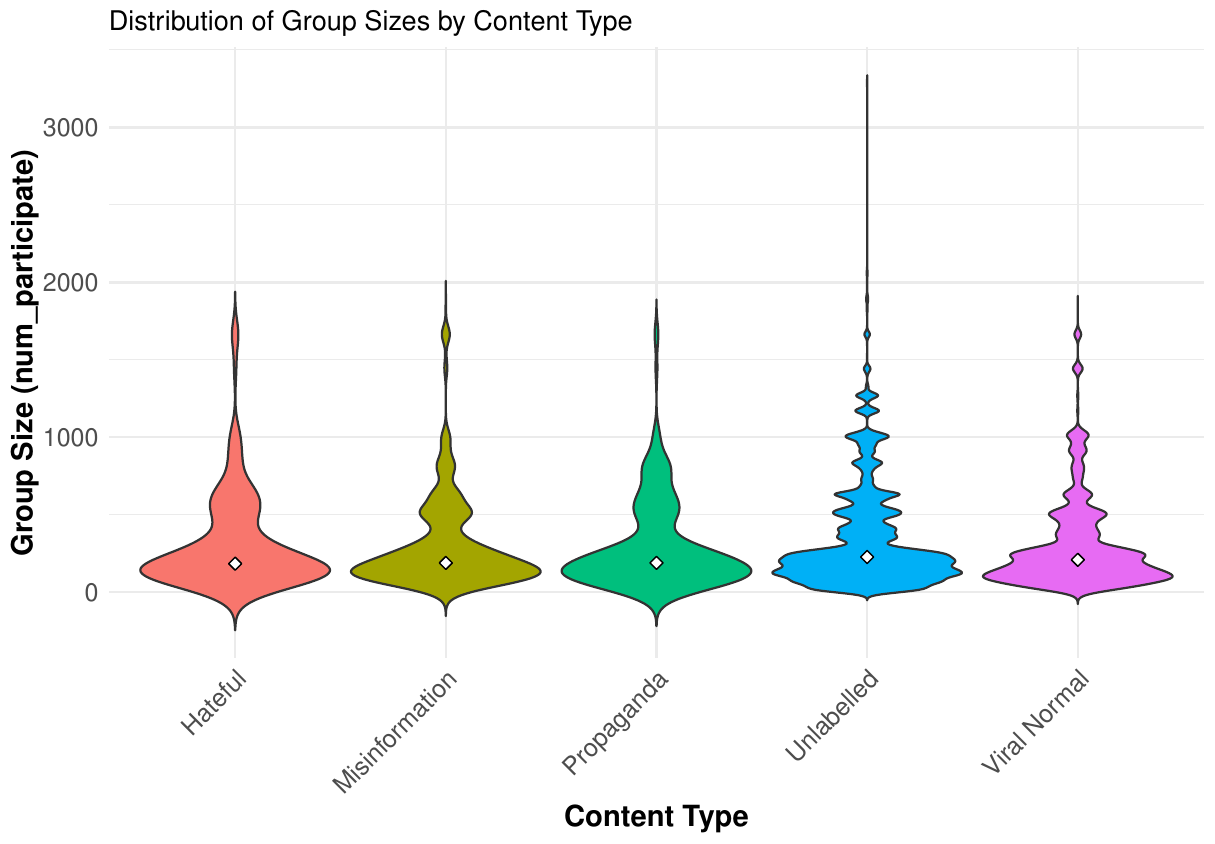}
    \caption{Violin plot of group size distribution by content type. No substantial difference in group sizes is observed across content types. Wider tails in the \texttt{unlabeled} category are likely due to its larger volume.}
    \label{fig:violin_group_size_content_type}
\end{figure}

To assess whether content type is systematically associated with different group sizes, we examine the distribution of group sizes by content type using violin plots. As shown in Figure~\ref{fig:violin_group_size_content_type}, there is no clear systematic difference in group size across content types. The group size distributions are largely overlapping, suggesting that differences in estimated population reach are not merely driven by group size heterogeneity across content types. While the unlabeled category exhibits more extreme large-group outliers, this is likely due to the greater volume of such messages in our dataset, rather than a structural association between content type and group size.

We then combine the empirical group size distribution for different content types with the estimated structural parameters $b$ (breadth) and $h$ (depth) from the influence cascade model to regenerate synthetic cascades and assign group sizes accordingly. This enables us to approximate the total number of individuals potentially exposed to each type of message.

\begin{figure}[ht]
    \centering
    \includegraphics[width=0.5\textwidth]{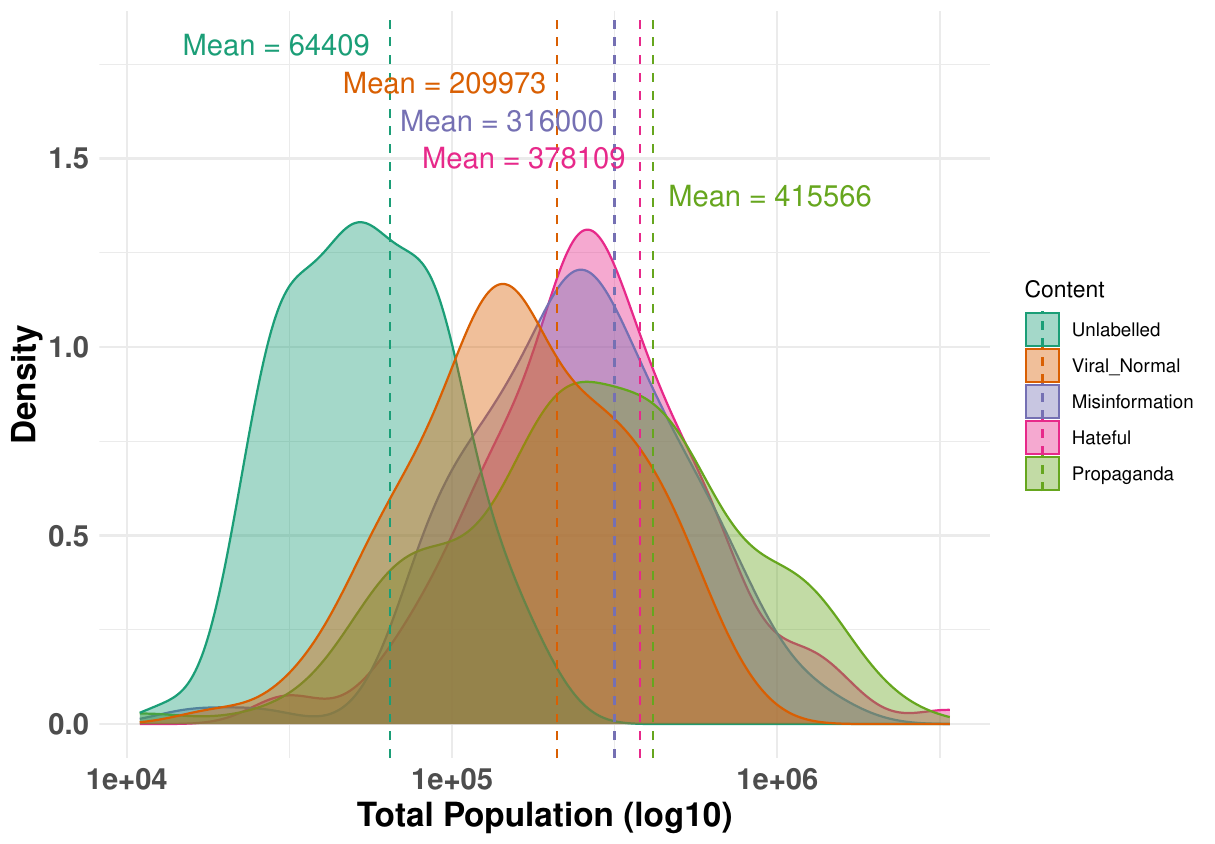}
    \caption{Density of influenced population by content type based on influence cascade. Vertical dashed lines indicate the mean influenced population for each content type.}
    \label{fig:influence_cascade_influenced_population_plot}
\end{figure}

Figure~\ref{fig:influence_cascade_influenced_population_plot} presents the distribution of population-level impact estimates by content type. Our findings suggest that harmful messages, due to their larger breadth and depth, tend to reach significantly more people than normal messages. According to our estimates, harmful messages affect approximately five times more individuals than normal messages. Even when focusing on normal messages with a forwarding score greater than or equal to five, harmful messages still reach nearly twice the audience. Moreover, we observe differences among harmful message categories: propaganda and hateful speech tend to have the highest reach, while the impact of misinformation is relatively lower.


\subsection{Demographic Analysis of Cascade Structures}
In this section, we build on the conclusions of this study by incorporating demographic features to further investigate the structural properties of cascades. For each cascade, we analyzed the demographic characteristics of the individual who first initiated the message, with the aim of determining whether there are structural differences in cascades started by people with varying demographic backgrounds. Specifically, we considered five demographic features: caste, religion, income, education, and age. Detailed results are provided in the appendix.

As shown in Table \ref{tab:cascade_demor}, we did not find any significant structural differences in the cascades. Although this suggests that demographic factors may not have a substantial impact on the breadth or depth of information cascades, further analysis is required to draw definitive conclusions. Future studies could explore these factors more thoroughly to understand whether subtle or context-specific effects could exist.

\begin{table}[h]
  \centering
  \caption{Comparison of Structural Parameters ($\mu_b$, $\sigma_b$, $\mu_h$, $\sigma_h$) for Influence Cascades Across Demographic Features}
  \label{tab:cascade_demor}
  \begin{tabular}{llcccc}
    \toprule
 Feature & Category & $\mu_b$ & $\sigma_b$ & $\mu_h$ & $\sigma_h$ \\
    \midrule
    Income & $<$25k       & 2.79 & 0.382 & 4.47 & 0.164 \\
           & $\geq$25k    & 2.73 & 0.292 & 4.44 & 0.126 \\
    \midrule
    Religion & Hinduism   & 2.78 & 0.380 & 4.47 & 0.163 \\
             & Islam      & 2.75 & 0.330 & 4.46 & 0.141 \\
    \midrule
    Caste & General       & 2.80 & 0.390 & 4.48 & 0.167 \\
          & Other         & 2.76 & 0.352 & 4.46 & 0.151 \\
    \midrule
    Education & $<$High school    & 2.77 & 0.366 & 4.46 & 0.157 \\
              & $\geq$High school & 2.79 & 0.371 & 4.47 & 0.160 \\
    \midrule
    Age & $<$25          & 2.80 & 0.368 & 4.48 & 0.158 \\
         & 25–34         & 2.84 & 0.425 & 4.49 & 0.182 \\
         & $>$35         & 2.73 & 0.344 & 4.45 & 0.148 \\
    \bottomrule
  \end{tabular}
\end{table}

\section{Discussion}

In this paper, we study the structural dynamics of harmful content dissemination in WhatsApp groups. We construct a new dataset through data donations, covering 5,953 groups in India and consisting of 5,158,879 messages that span text, images, and videos. Using this large-scale dataset, we apply an algorithm developed by \citet{Maunuel2012} to reconstruct the dissemination paths between the groups in our sampled data. We then use the algorithm developed by \citet{Eldar2011} to estimate two key structural parameters of the complete information diffusion process from partially observed cascades in our sample: breadth and depth. Our study reveals several key findings: First, harmful messages, such as misinformation, hate speech, and propaganda, tend to have significantly greater breadth and depth of dissemination compared to normal messages. Furthermore, propaganda exhibits a wider and deeper spread compared to misinformation. We also found that the dissemination of harmful messages is primarily driven through video and image formats. However, differences in message modality alone are insufficient to fully explain the significant structural differences in the dissemination processes between harmful and normal messages. Finally, we estimate the population-level impact of harmful messages, finding that, on average, harmful messages affect approximately five times more people than normal messages.

To our knowledge, this is the first effort to build a new dataset on private groups based on data donation. However, we also acknowledge certain limitations in the methodology used in this paper. First, the second algorithm developed by \cite{Eldar2011} is based on several strong assumptions, such as uniform sampling. In future work, our aim is to extend this model to obtain more accurate population-level estimates. In addition, we intend to further investigate the underlying factors that contribute to the structural differences between harmful and normal messages. Finally, we hope to integrate demographic data to explore how harmful messages affect different demographic groups differently.

\paragraph{Code and Data Availability}
The analysis code used to produce the results and figures is available at \url{https://github.com/YuxinLiu1997/Structural-Dynamics-of-Harmful-Content-Dissemination-on-WhatsApp}. Due to privacy protections and data-sharing agreements, the WhatsApp dataset used in this study cannot be made publicly available.

\clearpage

\appendix

\section{Validation of Cascade Reconstruction Method Using Forwarding Scores}

In Table \ref{tab:cascade_forwarding}, we categorize the data based on the forwarding score, a feature of WhatsApp that records the number of times a message is forwarded. When a message is forwarded more than five times, it is labeled "forwarded many times." Using this characteristic, we reclassify the cascade dataset and apply the same reconstruction algorithm to datasets with different forwarding scores. We expect that as the forwarding score increases, the corresponding parameters, breadth (\(b\)) and depth (\(h\)), will also increase. This would further validate the accuracy of our method.
\begin{table}[h]
  \centering
  \caption{Comparison Between Different Forwarding Scores for Influence Cascades}
  \label{tab:cascade_forwarding}
  \begin{tabular}{lcccc}
    \toprule
    F.S. & $\mu_b$ & $\sigma_b$ & $\mu_h$ & $\sigma_h$ \\
    \midrule
    0 & 2.79 & 0.326 & 4.47 & 0.140 \\
    1 & 2.78 & 0.292 & 4.47 & 0.125 \\
    2 & 2.83 & 0.301 & 4.49 & 0.129 \\
    3 & 2.92 & 0.421 & 4.53 & 0.180 \\
    4 & 3.05 & 0.538 & 4.58 & 0.230 \\
    $\geq 5$ & 3.38 & 0.550 & 4.73 & 0.237 \\
    \bottomrule
  \end{tabular}
\end{table}


As shown in Table \ref{tab:cascade_forwarding}, both the breadth and depth increase as the forwarding score increases. In particular, as the breadth and depth grow, the scale of dissemination tends to increase exponentially. Therefore, a higher forwarding score indicates a wider and more extensive spread of the message. This observation provides further validation of the effectiveness of our method. Although the forwarding score was not explicitly factored into our estimation process, the estimated values of breadth and depth still successfully capture the characteristic that messages forwarded more frequently tend to propagate more widely and deeply. This consistency suggests that our method effectively reflects the structural dynamics of message dissemination, thereby enhancing its robustness and reliability in real-world scenarios.

\section{Robustness of Structural Differences Between Harmful and Normal Messages at different sampling rates}

To validate the robustness of our cascade reconstruction approach, we replicate the analysis using different sampling rates from $p = 0.01$ to $p=0.05$, compared to $p = 0.02$ used in the main results. The summary statistics of the estimated structural parameters—breadth ($b$) and depth ($h$)—are shown in Table~\ref{tab:influence_cascade_different_p}.

\begin{table}[h]
  \centering
  \caption{Estimated Structural Parameters Across Sampling Rates $p$}
  \label{tab:influence_cascade_different_p}
  \begin{tabular}{llcccc}
    \toprule
    $p$ & Content Type & $\mu_b$ & $\sigma_b$ & $\mu_h$ & $\sigma_h$ \\
    \midrule
    0.01 & Hateful        & 4.34 & 0.644 & 4.95 & 0.219 \\
         & Misinformation & 4.24 & 0.665 & 4.91 & 0.223 \\
         & Propaganda     & 4.40 & 0.765 & 4.96 & 0.253 \\
         & Viral Normal   & 4.00 & 0.702 & 4.83 & 0.234 \\
         & Unlabeled      & 3.27 & 0.390 & 4.58 & 0.132 \\
    \midrule
    0.03 & Hateful        & 3.49 & 0.517 & 4.86 & 0.286 \\
         & Misinformation & 3.41 & 0.536 & 4.81 & 0.290 \\
         & Propaganda     & 3.53 & 0.620 & 4.87 & 0.323 \\
         & Viral Normal   & 3.21 & 0.572 & 4.71 & 0.302 \\
         & Unlabeled      & 2.63 & 0.360 & 4.40 & 0.188 \\
    \midrule
    0.04 & Hateful        & 3.33 & 0.501 & 4.78 & 0.301 \\
         & Misinformation & 3.25 & 0.513 & 4.74 & 0.312 \\
         & Propaganda     & 3.37 & 0.585 & 4.81 & 0.356 \\
         & Viral Normal   & 3.06 & 0.549 & 4.63 & 0.325 \\
         & Unlabeled      & 2.49 & 0.358 & 4.31 & 0.189 \\
    \midrule
    0.05 & Hateful        & 3.21 & 0.473 & 4.73 & 0.332 \\
         & Misinformation & 3.13 & 0.489 & 4.68 & 0.339 \\
         & Propaganda     & 3.23 & 0.552 & 4.77 & 0.391 \\
         & Viral Normal   & 2.95 & 0.531 & 4.57 & 0.344 \\
         & Unlabeled      & 2.38 & 0.366 & 4.25 & 0.177 \\
    \bottomrule
  \end{tabular}
\end{table}

We observe that the structural differences between harmful and normal messages persist under different sampling rates. Specifically, harmful messages—including \textit{hateful}, \textit{misinformation}, and \textit{propaganda}—consistently exhibit higher mean breadth ($\mu_b$) and depth ($\mu_h$) compared to normal messages. For instance, the mean breadth for \textit{hateful} messages is $4.34$, significantly greater than $3.27$ for normal messages under sampling rate $p=0.1$. Even when controlling for forwarding score (i.e., comparing to Normal ($F.S.\geq 5$)), harmful messages still demonstrate greater structural spread, with \textit{propaganda} exhibiting a breadth of $4.40$, which exceeds the breadth of viral normal messages ($\mu_b = 3.89$) even when forwarding score is high and sampling rate is $0.01$.

These findings confirm that our conclusions about the distinct propagation characteristics of harmful messages are robust to different levels of sampling and not merely an artifact of forwarding score differences.

\section{Validation of Cascade Structure Robustness with Group Network}

\label{sec:group_overlap_validation}

\begin{table}[h]
  \centering
  \caption{Re-estimated Structural Parameters Based on Group Network with Overlap}
  \label{tab:group_overlap_robustness}
  \begin{tabular}{lcccc}
    \toprule
    Content Type & $\mu_b$ & $\sigma_b$ & $\mu_h$ & $\sigma_h$ \\
    \midrule
    Hateful         & 3.77 & 0.557 & 4.91 & 0.257 \\
    Misinformation  & 3.68 & 0.577 & 4.87 & 0.261 \\
    Propaganda      & 3.82 & 0.672 & 4.92 & 0.288 \\
    Viral Normal    & 3.48 & 0.618 & 4.77 & 0.269 \\
    Unlabeled       & 2.85 & 0.384 & 4.50 & 0.168 \\
    \bottomrule
  \end{tabular}
\end{table}

To address the concern that some users may belong to multiple groups—thus potentially connecting otherwise disjoint dissemination paths—we construct an alternative group-level network, where an edge is added between two groups if they share at least one common member.

Using this augmented network, we reconstructed the cascades and re-estimated the structural parameters \( b \) (breadth) and \( h \) (depth) for each content type. As shown in Table~\ref{tab:group_overlap_robustness}, the estimated values remain consistent with those reported in the main analysis. This suggests that the conclusions regarding the differences in virality across content types are robust to the inclusion of inter-group connectivity induced by shared users.

\end{document}